\documentclass[preprint,3p,times,sort&compress,fleqn]{elsarticle} 

\usepackage{amssymb}   
\usepackage{amsthm}    
\usepackage{amsmath}
\usepackage{booktabs}  
\usepackage{makecell}  
\usepackage[table]{xcolor}   
\usepackage{microtype}
\PassOptionsToPackage{hyphens}{url}
\usepackage{hyperref}  
\hypersetup{
    colorlinks,
    linkcolor={red!50!black},
    citecolor={blue!50!black},
    urlcolor={blue!80!black}
}
\newcommand{\details}[1]{{\scriptsize#1}}
\newcolumntype{L}{>{\small}l}
\newcolumntype{R}{>{\small}r}
\newcommand{\makedoublecell}[2]{\makecell[l]{#1 \\ #2}}
\newcommand{\greyrule}{\arrayrulecolor{black!25}\midrule[0.25pt]\arrayrulecolor{black}}
\def\bx{\mathbf{x}}
\def\bX{\mathbf{X}}
\usepackage[most]{tcolorbox}
\newtcolorbox{myhighlight}[1][]{%
    colback=black!5,
    colframe=black!5,
    notitle,
    sharp corners,
    borderline west={2pt}{0pt}{red!80!black},
    enhanced,
}
    
\journal{Information Fusion}

\begin{document}

\begin{frontmatter}

\title{Adversarial attacks and defenses in explainable artificial intelligence: A survey}

\author[uw]{Hubert Baniecki}\ead{h.baniecki@uw.edu.pl}
\author[uw,pw]{Przemyslaw Biecek}\ead{p.biecek@uw.edu.pl}

\affiliation[uw]{organization={University of Warsaw}, city={Warsaw}, country={Poland}}
\affiliation[pw]{organization={Warsaw University of Technology}, city={Warsaw}, country={Poland}}

\begin{abstract}
Explainable artificial intelligence (XAI) methods are portrayed as a remedy for debugging and trusting statistical and deep learning models, as well as interpreting their predictions. However, recent advances in adversarial machine learning (AdvML) highlight the limitations and vulnerabilities of state-of-the-art explanation methods, putting their security and trustworthiness into question. The possibility of manipulating, fooling or fairwashing evidence of the model's reasoning has detrimental consequences when applied in high-stakes decision-making and knowledge discovery. This survey provides a comprehensive overview of research concerning adversarial attacks on explanations of machine learning models, as well as fairness metrics. We introduce a unified notation and taxonomy of methods facilitating a common ground for researchers and practitioners from the intersecting research fields of AdvML and XAI. We discuss how to defend against attacks and design robust interpretation methods. We contribute a list of existing insecurities in XAI and outline the emerging research directions in adversarial~XAI (AdvXAI). Future work should address improving explanation methods and evaluation protocols to take into account the reported safety issues.
\end{abstract}


\end{frontmatter}

\section{Introduction}\label{sec:introduction}

Explainable artificial intelligence (XAI) methods (for a brief overview see \citep{holzinger2022xai}, and for a comprehensive survey refer to \citep{schwalbe2023comprehensive}), e.g. post-hoc explanations like PDP~\citep{friedman2001greedy}, SG~\citep{simonyan2013deep}, LIME~\citep{ribeiro2016why}, IG~\citep{sundararajan2017axiomatic}, SHAP~\citep{lundberg2017unified}, TCAV~\citep{kim2018interpretability}, Grad-CAM~\citep{selvaraju2020gradcam} to name a few, provide various mechanisms to interpret predictions of machine learning models. A popular critique of XAI, in favour of inherently interpretable models, is its inability to faithfully explain the black-box predictive function~\citep{rudin2019stop}. Nevertheless, explanations find success in applications like autonomous driving~\citep{gu2020certified} or drug discovery~\citep{jimenez2020drug}, and can be used to better understand the reasoning of large models like AlphaZero~\citep{mcgrath2022acquisition}. 

\begin{figure}[t]
    \centering
    \includegraphics[width=0.99\linewidth]{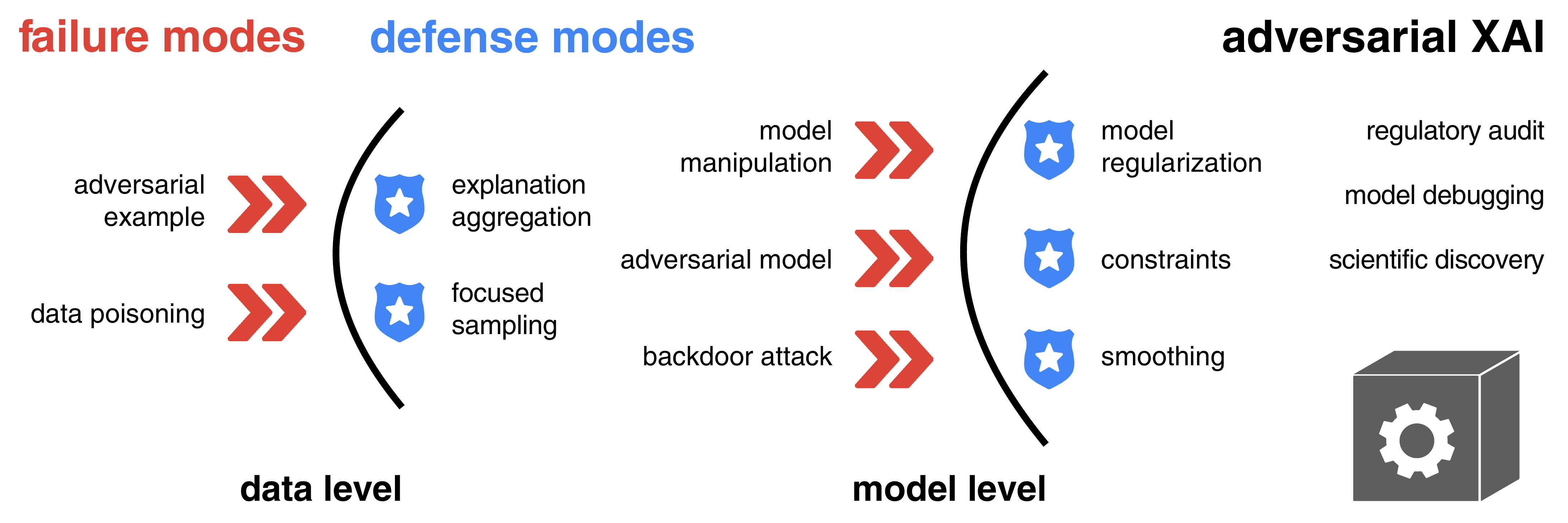}
    \caption{Explainable AI methods have vurlnerabilities related to safety and security, which we call \emph{failure modes}. They mainly exist on two levels exploiting manipulation of data or the model. Adversarial manipulation of explanations leads to misinterpretation of the model's behaviour in audits, debugging and scientific discovery. The most common adversarial attacks threatening current machine learning explanations are adversarial examples, data poisoning, model manipulation, and backdoors. There exist defense mechanisms like explanation aggregation and model regularization that aim to improve the robustness of XAI applications.}
    \label{fig:summary}
\end{figure}

Recently, studies related to adversarial machine learning \citep[AdvML,][]{kolter2018adversarial, rosenberg2021adversarial, machado2021adversarial} became more prevalent in research on XAI, which uncovered the vulnerabilities of explanation methods that raise concerns about their trustworthiness and security~\citep{papernot2018sok}. Adverarial attacks like data poisoning~\citep{zhang2021data,baniecki2022manipulating,brown2022making,laberge2023fooling}, model manipulation~\citep{heo2019fooling,dimanov2020you,anders2020fairwashing,slack2020fooling} and backdoors~\citep{viering2019how,noppel2023disguising} become prominent \emph{failure modes} of XAI methods. Simultaneously, defense modes like focused sampling of data~\citep{ghalebikesabi2021locality,vres2022preventing} and model regularization~\citep{chen2019robust,boopathy2020proper,wang2020smoothed,dombrowski2022towards} are proposed to counteract these attacks. Figure~\ref{fig:summary} illustrates common failure modes that exist due to data and models influencing the outputs of explanation methods. The main goal of an attacker is to manipulate the model's explanation by changing data or a model so as to deceive the explanations' recipient~\citep{poursabzi2021manipulating}. It becomes a real threat in sensitive applications like explaining decision making systems that help judges make bail decisions~\citep{lakkaraju2020how}. Naturally, attacks on explanations vary based on the considered machine learning task, model architecture, and class of XAI methods, which is not well systemized.

Figure~\ref{fig:adversarial-example} presents the most common of such attack mechanisms, referred to as \emph{adversarial example}~\citep{biggio2013evasion,szegedy2014intriguing,ghorbani2019interpretation,kindermans2019unreliability,dombrowski2019explanations}, i.e. a slight perturbation applied to an input image drastically changes the explanation of the ``dog'' class predicted by a model. It turns out that having access to the model, one can use differentiation to optimize input perturbations that extensively affect the explanations without impacting the model's predictive performance. In a more advanced setting, it is possible to poison the training data used for model training, so as to manipulate explanations only for a set of pre-defined inputs, e.g. coming from a particular class. Over the time, alternative optimization methods that do not require access to the model were proposed to perform such manipulation~\citep{sinha2021perturbing,huang2023focusshifting}.

In Figure~\ref{fig:adversarial-example}, we can also observe a simple defense mode that counteracts adversarial examples. An aggregation of explanations obtained with different algorithms shows to be less susceptible to such input manipulation~\citep{rieger2020simple}. This is because the attacker targeted a single explanation method and their ensemble is more robust. There is an open debate whether the most effective defense mechanism is to patch explanation methods~\citep{ghalebikesabi2021locality,blesch2023unfooling} or improve model algorithms, e.g. with regularization in training~\citep{chen2019robust,wicker2023robust}.

\paragraph{Related surveys} There certainly exists broad literature surveying explanation methods~\citep{guidotti2018survey,schwalbe2023comprehensive}, adversarial attacks on model predictions~\citep{machado2021adversarial,cina2023wild}, and specifically the application of XAI in AdvML~\citep{liu2021adversarial}. However, community lacks a comprehensive overview of research concerning adversarial attacks on explanations and fairness metrics, which would allow to set common ground for future work. Notably, \citet{mishra2021survey} briefly review the robustness of feature importance and counterfactual explanations. It remains important to constantly track the appearing failure modes of XAI and patch the explanation frameworks.

\paragraph{Contribution} In this survey, we aim to highlight the rapidly emerging cross-domain research in what we call adversarial explainable AI (AdvXAI). Assessing the scope of security threats in XAI is key to the successful adoption of explanation methods in practical applications. We contribute a systemization of the current state of knowledge concerning \emph{adversarial attacks on model explanations}~(Section~\ref{sec:attacks}) and \emph{defense mechanisms against these attacks} (Section~\ref{sec:defense}). We moreover confront it with the closely related work concerning \emph{attacks on machine learning fairness metrics}~(Section~\ref{sec:fairness}). A~thorough overview of over 50 papers allows us to specify the opportunities, challenges and future research directions in~AdvXAI~(Section~\ref{sec:discussion}). 

This review serves as an approachable outlook to recognize the potential research gaps and define future directions. We first included visible papers published in major machine learning conferences~(ICML, ICLR, NeurIPS, AAAI) and journals~(AIj, NMI) since~\citet{ghorbani2019interpretation}. We then extensively searched their citation networks for papers related to AdvXAI published in other venues. We purposely exclude a large number of papers focusing primarily on explanation evaluation without relating to the adversarial scenario~\citep[refer to][]{vilone2021notions,nauta2023from}.

\section{Background}

Here, we provide a brief introduction to methods of adversarial attacks on machine learning predictions (Section~\ref{subsec:advml}), as well as methods used for explaining artificial intelligence models (Section~\ref{subsec:xai}). Readers familiar with these basic concepts can skip to Section \ref{sec:attacks}. 

\subsection{Notation}

Throughout this paper, we mainly consider a supervised machine learning task where model $f_{\theta}: \mathcal{X} \mapsto \mathcal{Y}$ is trained to predict an output feature $\mathbf{Y} \in \mathcal{Y} \subseteq \mathbb{R}$ using input features $\bX \in \mathcal{X} \subseteq \mathbb{R}^{d}$. Here, $\theta$ represents the model's parameters, e.g. weights of a neural network. We will write it simply as $f$ when no confusion can arise. Let $\bx \in \mathcal{X}$ be an input vector for which the prediction $f(\bx)$ we want to explain. Further, let $g(\cdot{,}\cdot)$ denote an explanation function, for which both model and data are the input, where the output domain varies between different explanation methods. For example, $g(f_{\theta},\cdot): \mathcal{X} \mapsto \mathbb{R}^d$ maps each input feature to its importance or attribution to the model's prediction. In principle, $g(f,\bx)$ denotes a \emph{local explanation} corresponding to a particular prediction, e.g. gradient $g(f,\bx) = \nabla_{\bx} f(\bx)$, while $g(f,\bX)$ denotes a \emph{global explanation}, e.g. feature importance for tabular data. Our goal is to introduce a simple unified notation facilitating a common ground for researchers and practitioners from the intersecting research fields of AdvML and XAI. In what follows, the symbol $\rightarrow$ denotes an adversarial change in a given object, e.g. a small perturbation of input becomes $\bx \rightarrow \bx'$. We will use the symbol $\approx$ to denote similarity between two values, e.g. similar predictions $f(\bx) \approx f(\bx')$, and contrary $\neq$ to denote dissimilarity, e.g. visibly different explanations $g(f,\bx) \neq g(f,\mathbf{x'})$.

\begin{figure}[t]
    \centering
    \includegraphics[width=0.33\linewidth]{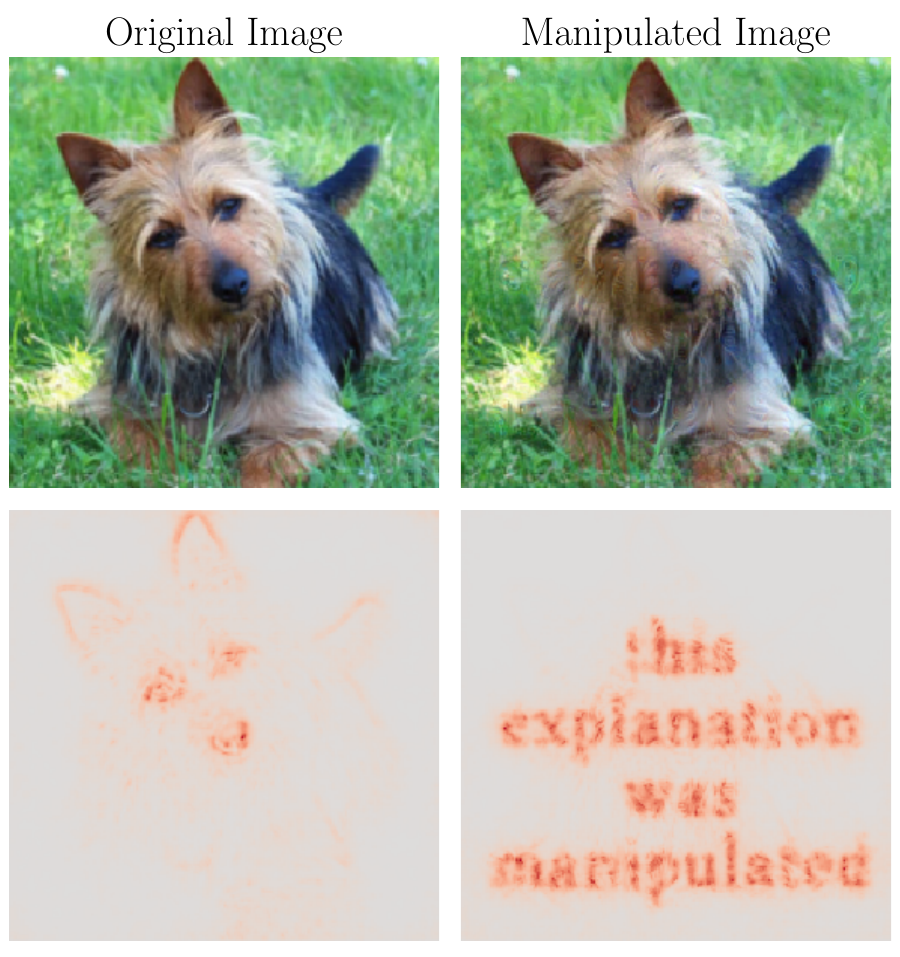}
    \includegraphics[width=0.66\linewidth]{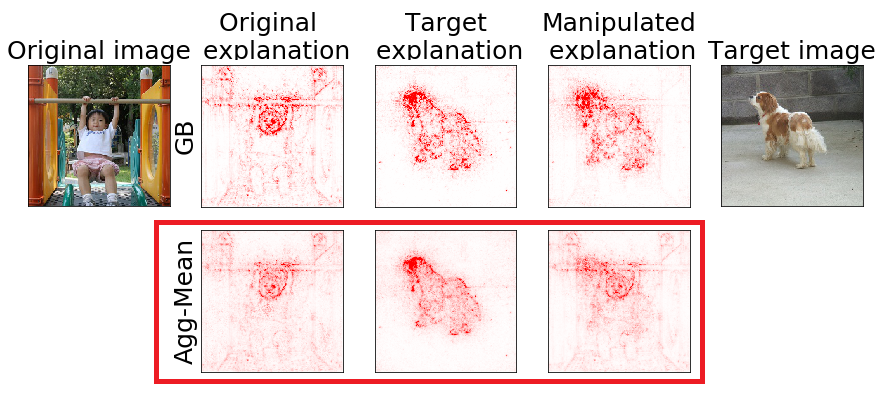}
    \caption{Adversarial example is the most common attack on local explanations of the image classifier's prediction. \emph{Left} \citep[adapted from][]{dombrowski2019explanations}: An original image is classified as a ``dog'' and its explanation points out to features influencing this decision. The image can be adversarially changed with perturbation that remains invisible to human eye. The manipulated image remains classified as a ``dog'', but its explanation drastically changed.
    \emph{Right} \citep[adapted from][]{rieger2020simple}: It is possible to defend from adversarial examples by ensembling explanations resulting from multiple algorithms. In this example, an attacker aims to change the guided backprop explanation for an original image so that it looks similar to the explanation for a target image; without impacting the model's prediction. While the guided backprop explanation is successfully manipulated, an aggregated mean of outputs from three alternative explanation methods remains more resistant to adversarial attacks.}
    \label{fig:adversarial-example}
\end{figure}

\subsection{Adversarial machine learning}\label{subsec:advml}

Since 2004, there has been an enormous development of adversarial methods in (deep) machine learning \citep{biggio2018wild}. In general, these are algorithms that aim to attack the model's behaviour, which relates to the security, safety, and robustness of AI systems. The most explored class of attacks in machine learning is an \emph{adversarial example} \citep{biggio2013evasion,szegedy2014intriguing,papernot2016limitations}, especially in computer vision tasks, where one aims to minimally modify input data so that it fools a model into misclassification.\footnote{Adversarial examples are closely related to counterfactual explanations where the latter aim to interpret the model instead of deceiving it~\citep{ignatiev2019relating}.} We use the introduced notation to describe this attack strategy in the following way:
\begin{equation}
    \bx \rightarrow \bx'
        \Longrightarrow 
    f(\bx) \neq f(\bx').
\end{equation}
Adversarial examples are an orthogonal concept to fooling examples -- random inputs, e.g. images unrecognizable by humans, for which models predict with very high confidence~\citep{nguyen2015deep}. \citet{brown2017adversarial} propose a method to create adversarial patches that trigger a misclassification when added to an input. Consecutively, \citet{athalye2018synthesizing} demonstrate the existence of 3D adversarial objects in the physical world, which are misclassified when photographed from multiple viewpoints.
In their seminal paper, \citet{su2019one} show that it is possible to fool a model into misclassification by perturbing only a single pixel in an image. While imaging data may be the most illustrative, adversarial examples were originally studied for machine learning models predicting on tabular and text data \citep{fumera2006spam,biggio2013evasion}. Adversarial perturbations of text inputs include misspellings, punctuation errors, and swapping words or characters~\citep{rychalska2019models}.

The threat of manipulating model predictions with adversarial inputs increases the risk of deploying and applying AI systems in the real world. Algorithms that aim to counter attacks become defenses, which include data augmentation~\citep{machado2021adversarial}, model regularization~\citep{gu2015towards}, and distillation~\citep{papernot2016distillation}. Their ultimate goal is to make it harder for an attacker to fool the model's behaviour, i.e. minimize the potential risk. Often, detecting the attack is a preliminary (and sufficient) step in a successful defense~\citep{metzen2017detecting}.

Another widely studied failure mode of machine learning models is a \emph{backdoor attack} \citep{gu2017badnets,chen2017targeted}, in which (generally) one assumes that the adversary has access to the model, e.g. in a scenario of outsourced training. Its goal is to make the model predict wrong for new inputs with a specific trigger pattern known to an adversary. The most common approach to creating a backdoor is to \emph{poison data} used for training in a way so that the adversarial model remains indistinguishable from the desired one:
\begin{equation}
    \left.\begin{array}{@{}r@{}}
        \bX \rightarrow \bX' \Rightarrow f_{\theta} \rightarrow f_{\theta'} \\
        \bx \rightarrow \bx' \\
    \end{array}\right\}
        \Longrightarrow 
        f_{\theta}(\bx) \approx f_{\theta'}(\bx) \neq f_{\theta'}(\bx').
\end{equation}
More broadly, various poisoning attacks on machine learning models have been proposed targeting different adversarial goals~\citep{tian2022comprehensive}, e.g. decreasing classification accuracy~\citep{biggio2012poisoning}. \citet{cina2023wild} introduce a comprehensive systematization of poisoning attacks (and defenses) related to model predictions. Attacks are categorized based on the attacker's goal, e.g. violations of availability or integrity; knowledge, e.g. having complete access to the system or treating it as a black box; capability, e.g. possible ways of perturbing data; and lastly attacker's strategy, e.g. perturbing training or test data. Defenses are categorized based on whether they sanitize data or modify the learning algorithm/model. Such systematization of adversarial attacks facilitates further research on AI insecurities.

\subsection{Explainable artificial intelligence}\label{subsec:xai}

In parallel with research on AdvML, there has been an increasing interest in methods that allow us to understand and interpret machine learning models~\citep{miller2019explanation,biecek2021explanatory}. XAI (or explanation) methods become useful in various applications like regulatory audit~\citep{krishna2023bridging}, model debugging~\citep{hryniewska2021checklist}, or scientific discovery~\cite{jimenez2020drug}. The origins of explainability relate to AI systems that should provide reasoning for their decisions~\citep{amgoud2009using,amgoud2022axiomatic}, often comprehensible by humans. Currently, there is an open discussion on whether the developed XAI methodology is sufficient to attain trustworthy AI systems~\citep{miller2023explainable,ali2023explainable}.

For example, consider a simple gradient of the (differentiable) model's output with respect to input as an intuitive explanation of feature influence~\citep{baehrens2010how,simonyan2013deep}:
\begin{equation}
    g(f,\bx) = \nabla_{\bx} f(\bx).
\end{equation}
The most explored class of explanations in machine learning are local post-hoc \emph{feature attributions}~\citep{vstrumbelj2010efficient,ribeiro2016why,lundberg2017unified,sundararajan2017axiomatic}, which indicate how much each input feature contributed to the model's outcome prediction:
\begin{equation}
    f(\bx) = \sum_{i = 1}^{d} g(f,\bx)_{i}.
\end{equation}
A state-of-the-art approach to obtaining vector $g(f,\bx)$ are algorithms estimating Shapley values~\citep{vstrumbelj2010efficient,lundberg2017unified,aas2021explaining}, a solution originating in game theory. Nowadays, there is an abundance of variations considering different data modalities, model algorithms, explanation quality metrics, or the desired computational efficiency~\citep{chen2023algorithms}. Specifically for tabular data, perturbation-based explanation methods~\citep{ribeiro2016why,lundberg2017unified} allow to explain individual predictions in a model-agnostic manner. Contrary, gradient-based~\citep{sundararajan2017axiomatic,lundstrom2022rigorous} and propagation-based~\citep{bach2015onpixelwise} local post-hoc explanations are more specific to deep neural networks, which are state-of-the-art predictive models trained on unstructured data, e.g. image and text.

Complementary to local explanations are global explanations that summarise patterns in model predictions consistent across a data distribution. Global \emph{feature importance}~\citep{breiman2001random,fisher2019all,molnar2023model} metrics quantify the model's reliance on a particular feature. For example, an aggregation of feature attributions for a subset of inputs becomes a natural importance metric~\citep{covert2020understanding}:
\begin{equation}
    G(f,\bX, g) = \sum_{\bx \in \bX} \lvert g(f,\bx) \rvert.
\end{equation}
\emph{Feature effect}~\citep{friedman2001greedy,moosbauer2021explaining,molnar2023model} explanations, e.g. partial dependence plots~\citep{friedman2001greedy}, visualise a global relationship between the expected model's prediction and values of a particular feature. Global explanations specific to deep neural networks include concept-based explanations~\citep{kim2018interpretability}, which relate human-understandable concepts to the predicted classes, e.g. how sensitive a prediction of ``zebra'' is to the presence of stripes in an image.

Note that there are various other approaches to explaining machine learning models~\citep[see e.g.][]{guidotti2018survey, holzinger2022xai, schwalbe2023comprehensive} and we referred to a non-exhaustive subset of explanation methods most targeted by adversarial attacks surveyed in this paper. Moreover, closely related to AdvXAI is research on evaluating the robustness of explanations~\citep{mishra2021survey} and applying explanation methods in adversarial scenarios~\citep{liu2021adversarial}.

\section{Adversarial attacks on model explanations}\label{sec:attacks}

To the best of our knowledge, \citet{ghorbani2019interpretation} is the first contribution to mention\footnote{Note that we acknowledge papers in order of date published, i.e. presented at a conference, as opposed to the first date appearing online, e.g. as a preprint or final proceedings version.} 
and propose an \emph{adversarial attack against explanation methods}, specifically gradient-based feature attributions~\citep{simonyan2013deep,sundararajan2017axiomatic} of (convolutional) neural networks. An intuition behind the attack is to iteratively perturb an input in the gradient direction of explanation change: $\bx_{i+1} = \bx_{i} + \alpha \cdot \text{sign}\big(\nabla_{\bx} g\left(f, \bx_{i}\right)\big)$. Given a number of iterations and step size $\alpha$, the algorithm results in a similar input $\bx'$ that potentially has a drastically different explanation $g(f, \bx')$. This optimization procedure is subject to constraints that the perturbation remains small $\|\bx - \bx'\|_{\infty} \leq \epsilon$ and both inputs have the same predicted class $f(\bx) = f(\bx')$. A similar strategy is used to attack another class of explanation methods -- influence functions~\citep{koh2017understanding}, which return the training inputs' importance on the loss of the explained input. Although experiments with image data highlight the vulnerability of explanation methods, this is fundamentally because the neural network model itself is vulnerable to such attacks~\citep{ghorbani2019interpretation}.

Previous related work discussed the worst-case (adversarial) notion of explanation robustness \citep[][page~7]{melis2018towards} and the notion of explanation sensitivity~\citep{ancona2018towards,kindermans2019unreliability}. Crucially, \citet{adebayo2018sanity} introduced randomization tests showing that a visual inspection of explanations alone can favour methods compelling to humans. It raised attention to the need for evaluating the explanations' quality, especially for deep models, with possible implications in adversarial settings. 

Table~\ref{tab:summary} introduces the notation we use to describe the 10 most common attacks on explanations of machine learning models. In Table~\ref{tab:attacks}, we summarize a comprehensive list of works that introduce attacks on explanation methods. We divide them based on the corresponding strategy of changing data, model, or changing both data and the model. For each of the attacks, we record the mentioned data modalities with the corresponding datasets used in experiments, as well as model algorithms. In general, current attack algorithms are divisible into those specific to neural networks~\citep{heo2019fooling} and model-agnostic attacks that can be applied to any black-box predictive function~\citep{slack2020fooling}, including neural networks.

\begin{table}[ht]
    \centering
    \caption{A summary of notation we use to describe the 10 most common attacks on explanations of machine learning models.}
    \label{tab:summary}
    \vspace{1em}
    \begin{tabular}{@{\hspace{1pt}}LlL@{\hspace{1pt}}}
        \toprule
            \textbf{Attack} & {\small\textbf{Notation}} & \textbf{References} \\
        \midrule
            adversarial example &  
            $
            \bx \rightarrow \bx'
                \Longrightarrow 
            \left\{\begin{array}{@{}l@{}}
                g(f,\bx) \neq g(f,\bx') \\ 
                f(\bx) \approx f(\bx') \\
            \end{array}\right.
            $ & \citep{ghorbani2019interpretation,kindermans2019unreliability,subramanya2019fooling,dombrowski2019explanations,kuppa2020blackbox,zhang2020interpretable,sinha2021perturbing,nanda2021fairness,huang2023safari} \\[1.25em]
            data poisoning (e.g. biased sampling) & 
            $
            \bX \rightarrow \bX'
                \Longrightarrow
            g(f,\bX) \neq g(f,\bX') 
            $ & \citep{fukuchi2020faking,baniecki2022fooling,brown2022making,baniecki2022manipulating,laberge2023fooling} \\[1.25em]
            data poisoning (in training) & 
            $
            \bX \rightarrow \bX'
                \Rightarrow 
            f_{\theta} \rightarrow f_{\theta'}
                \Longrightarrow
            g(f_{\theta},\bX) \neq g(f_{\theta'},\bX)
            $ & \citep{solans2020poisoning,mehrabi2021exacerbating,hussain2022adversarial} \\[1.25em]
            backdoor attack (fooling) &
            $
            \left.\begin{array}{@{}r@{}}
                \bX \rightarrow \bX' \Rightarrow f \rightarrow f' \\
                \bx \rightarrow \bx' \\
            \end{array}\right\}
                \Longrightarrow 
            \left\{\begin{array}{@{}l@{}}
                g(f,\bx) \neq g(f',\bx') \\  
                f'(\bx) \approx f'(\bx') \\
            \end{array}\right.
            $ & \citep{viering2019how,zhang2021data,noppel2023disguising} \\[1.25em]
            backdoor attack (red-herring) &
            $
            \left.\begin{array}{@{}r@{}}
                \bX \rightarrow \bX' \Rightarrow f \rightarrow f' \\
                \bx \rightarrow \bx' \\
            \end{array}\right\}
                \Longrightarrow 
            \left\{\begin{array}{@{}l@{}}
                g(f,\bx) \neq g(f',\bx') \\  
                f'(\bx) \neq f'(\bx') \\
            \end{array}\right.
            $ & \citep{noppel2023disguising} \\[1.25em]
            backdoor attack (full disguise) &
            $
            \left.\begin{array}{@{}r@{}}
                \bX \rightarrow \bX' \Rightarrow f \rightarrow f' \\
                \bx \rightarrow \bx' \\
            \end{array}\right\}
                \Longrightarrow 
            \left\{\begin{array}{@{}l@{}}
                g(f,\bx) \approx g(f',\bx') \\  f'(\bx) \neq f'(\bx') \\
            \end{array}\right.
            $ & \citep{noppel2023disguising} \\[1.25em]   
            model manipulation (e.g. fine-tuning) &
            $
            f_{\theta} \rightarrow f_{\theta'}
                \Longrightarrow 
            \left\{\begin{array}{@{}l@{}}
                \forall_{\bx \in \bX}\;g(f_{\theta},\bx) \neq g(f_{\theta'},\bx) \\ 
                \forall_{\bx \in \bX}\;f_{\theta}(\bx) \approx f_{\theta'}(\bx) \\
            \end{array}\right. 
            $ & \citep{heo2019fooling,dimanov2020you,anders2020fairwashing,slack2021counterfactual} \\[1.25em]  
            adversarial model (vs. local explanation) &
            $
            f \rightarrow f'
                \Longrightarrow 
            \left\{\begin{array}{@{}l@{}}
                \exists_{\bx \in \bX}\;g(f,\bx) \neq g(f',\bx) \\  
                \forall_{\bx \in \bX}\;f(\bx) \approx f'(\bx) \\
            \end{array}\right.
            $ & \citep{slack2020fooling,merrer2020remote} \\[1.25em]  
            adversarial model (vs. fairness metric) &
            $
            f \rightarrow f'
                \Longrightarrow 
            \left\{\begin{array}{@{}l@{}}
                g(f,\bX) \neq g(f',\bX) \\  
                \forall_{\bx \in \bX}\;f(\bx) \approx f'(\bx) \\
            \end{array}\right.
            $ & \citep{aivodji2019fairwashing,aivodji2021characterizing} \\[1.25em]
            trust manipulation &
            $
            g \rightarrow g'
                \Longrightarrow 
            \left\{\begin{array}{@{}l@{}}
                g(f,\bX) \neq g'(f,\bX) \\  
                \forall_{\bx \in \bX}\;g'(\bx) \approx f(\bx) \\
            \end{array}\right.
            $ & \citep{lakkaraju2020how} \\
         \bottomrule
    \end{tabular}
\end{table}

\paragraph{Adversarial example} Attacks relying on input perturbation change data to manipulate an explanation without impacting the model's prediction~\citep{ghorbani2019interpretation,kindermans2019unreliability,subramanya2019fooling,dombrowski2019explanations,kuppa2020blackbox,zhang2020interpretable}:
\begin{equation}
    \bx \rightarrow \bx'
        \Longrightarrow 
    \left\{\begin{array}{@{}l@{}}
        g(f,\bx) \neq g(f,\bx') \\ 
        f(\bx) \approx f(\bx') \\
    \end{array}\right..
\end{equation}

\paragraph{Model manipulation} Attacks relying on fine-tuning and weights regularization change the model to manipulate an explanation without impacting the model's predictive performance~\citep{heo2019fooling,dimanov2020you}: 
\begin{equation}
    f_{\theta} \rightarrow f_{\theta'}
        \Longrightarrow 
    \left\{\begin{array}{@{}l@{}}
        \forall_{\bx \in \bX}\;g(f_{\theta},\bx) \neq g(f_{\theta'},\bx) \\ 
        \forall_{\bx \in \bX}\;f_{\theta}(\bx) \approx f_{\theta'}(\bx) \\
    \end{array}\right..
\end{equation}

\begin{figure}[!ht]
    \centering
    \includegraphics[width=0.66\textwidth]{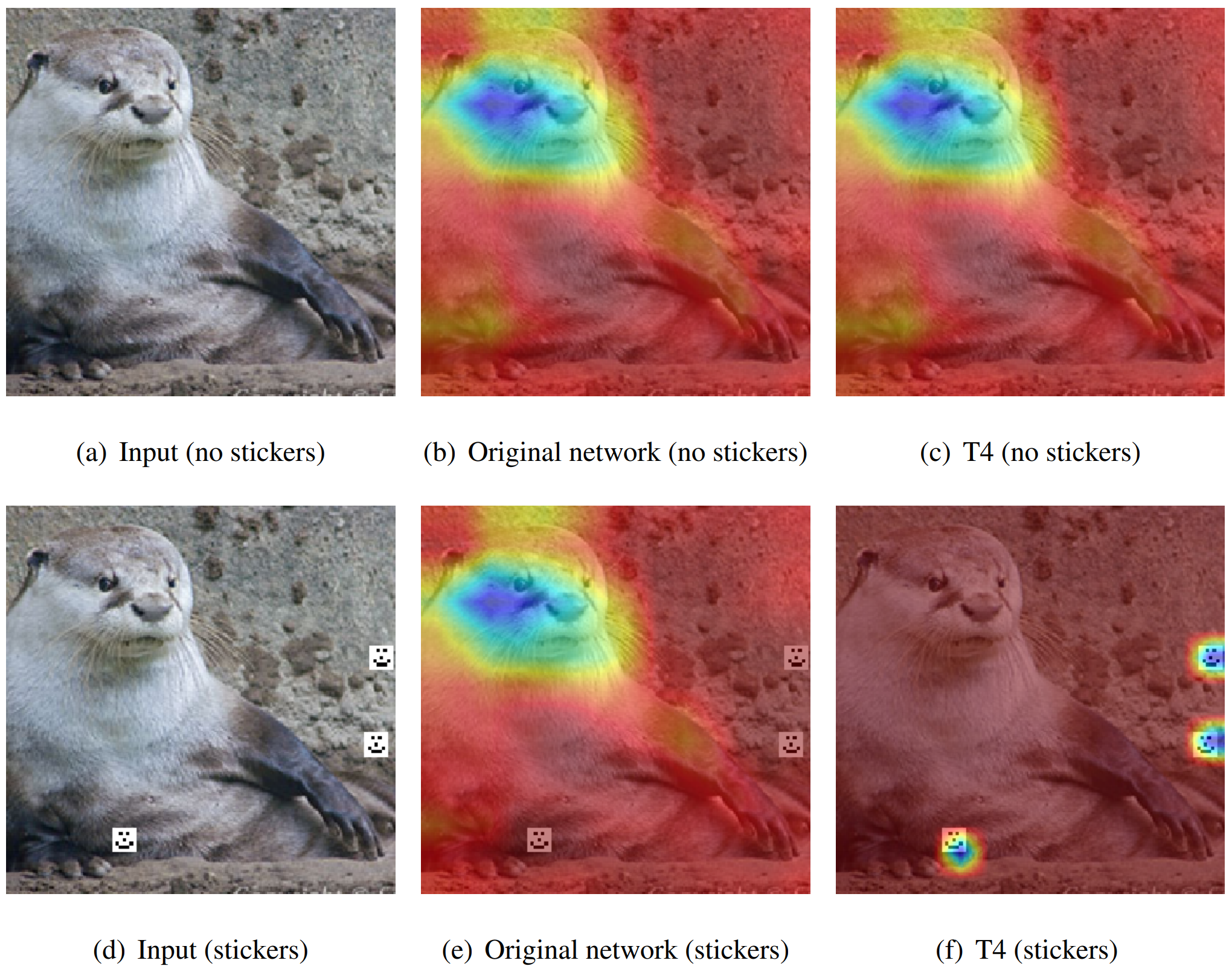}
    \caption{Illustrative example of a backdoor attack targeting the Grad-CAM explanation of an image classifier~\citep[adapted from][T4 means ``Technique~4'', i.e. a malicious explanation triggered by an input pattern]{viering2019how}. An attacker optimizes the model to manipulate explanations only for inputs with an added sticker, which is a specifically defined set of adversarial examples. The model with an encoded backdoor (T4) predicts the same class ``otter'' for both inputs, but its explanations drastically change when detecting the pattern (f).}
    \label{fig:backdoor}

    \vspace{1em}

    \centering
    \includegraphics[width=0.59\linewidth]{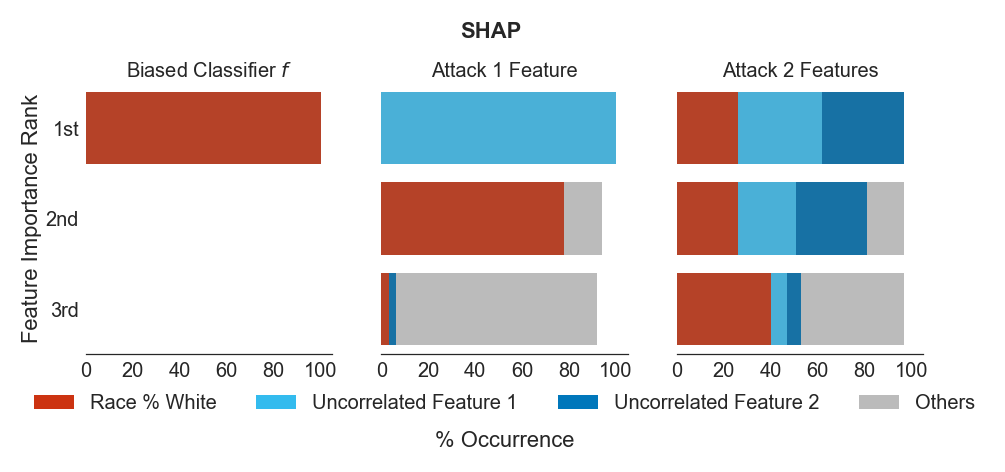}
    \includegraphics[width=0.40\linewidth]{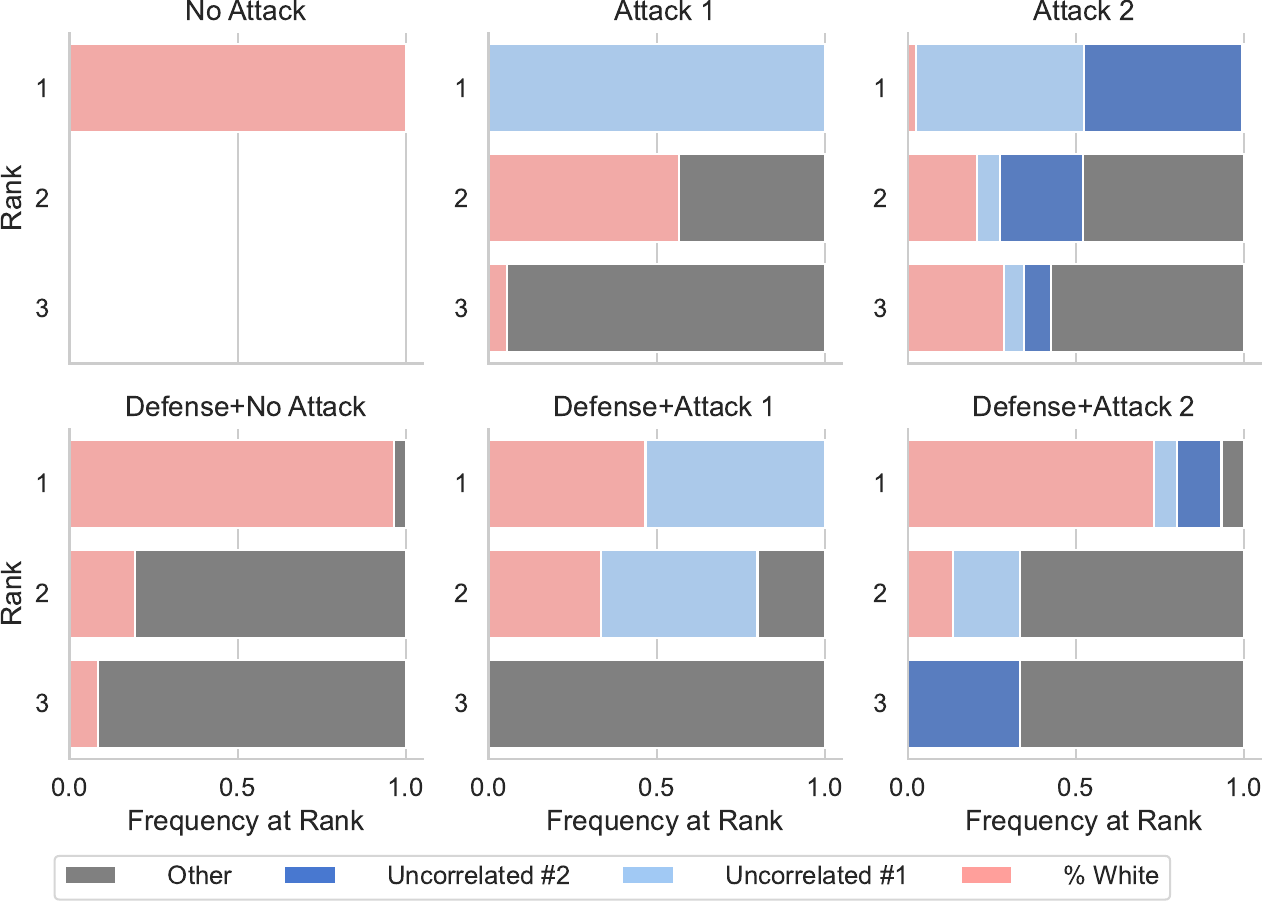}
    \caption{Deploying an adversarial model with manipulated explanations is a viable attack strategy to hide the bias of an unfair classifier predicting violent crime rate across various communities in the US. \emph{Left} \citep[adapted from][]{slack2020fooling}: A biased classifier relies on feature ``Race \% White'' to classify the outcome. An attacker uses a surrogate model that is unbiased out of data distribution to exploit the limitation of estimating SHAP -- relying on input perturbations. It is possible to manipulate an explanation so that it shows an uncorrelated feature (or two) being more important than ``Race \% White'', while simultaneously the black-box remains to predict unfairly in-distribution.
    \emph{Right} \citep[adapted from][]{carmichael2023unfooling}: Exemplary result of using a defending mechanism that partially uncovers an adversarial model attack. It imposes constraints on the data distribution used to estimate SHAP, which leads to retrieving an explanation similar to the the original one (``No Attack'') that shows the model is most influenced by ``Race \% White''.}
    \label{fig:fooling-shap}
\end{figure}

\paragraph{Backdoor attack} In general, it changes both data and a model as an input is modified to trigger a backdoor encoded in the model. The backdoor is most often included via poisoning the dataset used for model training~\citep{zhang2021data}: 
\begin{equation}
    \left.\begin{array}{@{}r@{}}
        \bX \rightarrow \bX' \Rightarrow f \rightarrow f' \\
        \bx \rightarrow \bx' \\
    \end{array}\right\}
        \Longrightarrow 
    \left\{\begin{array}{@{}l@{}}
        g(f,\bx) \neq g(f',\bx') \\  
        f'(\bx) \approx f'(\bx') \\
    \end{array}\right..
\end{equation}
\citet{viering2019how} manipulate Grad-CAM explanations of a convolutional neural network by changing its weights, but also proposes to leave a backdoor in the network (triggered by specific input patterns), which allows retrieving original explanations. Figure~\ref{fig:backdoor} shows an exemplary explanation resulting from a backdoor attack triggered by an input pattern. \citet{noppel2023disguising} extend fooling explanations through fine-tuning and backdoor to introduce so-called red-herring and fully disguising attacks. \emph{Red-herring} attack manipulates the explanation with the aim of covering an adversarial change in the model's prediction. For example, it can be applied to conceal adversarial examples that lead to misclassification:
\begin{equation}
    \left.\begin{array}{@{}r@{}}
        \bX \rightarrow \bX' \Rightarrow f \rightarrow f' \\
        \bx \rightarrow \bx' \\
    \end{array}\right\}
        \Longrightarrow 
    \left\{\begin{array}{@{}l@{}}
        g(f,\bx) \neq g(f',\bx') \\  
        f'(\bx) \neq f'(\bx') \\
    \end{array}\right..
\end{equation}
\emph{Fully-disguising} attack aims to show the original explanation for a changed prediction instead:
\begin{equation}
    \left.\begin{array}{@{}r@{}}
        \bX \rightarrow \bX' \Rightarrow f \rightarrow f' \\
        \bx \rightarrow \bx' \\
    \end{array}\right\}
        \Longrightarrow 
    \left\{\begin{array}{@{}l@{}}
        g(f,\bx) \approx g(f',\bx') \\  f'(\bx) \neq f'(\bx') \\
    \end{array}\right..
\end{equation}

\begin{table*}[t]
    \setlength{\tabcolsep}{3.5pt}
    \centering
    \caption{Summary of works introducing adversarial attacks on explanations of machine learning models. Most attacks change at least one of the data~(D), model~(M), or explanation~(E). Experiments consider data modalities including image~(I), tabular~(T), language~(Lg); and model algorithms specifically neural networks~(N), or black-box models~(B) including neural networks. Explanation methods are either local (L) or global~(G). \ref{app:abbr} lists other abbreviations.}
    \label{tab:attacks}
    \vspace{1em}
    \begin{tabular}{@{\hspace{1pt}}LLLLL@{\hspace{1pt}}}
        \toprule
        \textbf{Attack} & \textbf{Changes \details{strategy}} & \textbf{Modality \details{dataset}} & \textbf{Model \details{algorithm}} & \textbf{Explanation \details{method}}\\
        \midrule
        \citet{ghorbani2019interpretation} & D \details{adversarial example} & I \details{ImageNet, CIFAR-10} & N \details{SqueezeNet, InceptionNet} & L \details{SG, IG, DeepLIFT} \\
        \citet{kindermans2019unreliability} & D \details{adversarial example} & I \details{MNIST, ImageNet} & N \details{MLP, CNN, VGG} & L \details{SG, GI, IG, LRP, ..} \\
        \citet{viering2019how} & M \& D \details{backdoor attack} & I \details{ImageNet} & N \details{VGG} & L \details{Grad-CAM} \\
        \citet{subramanya2019fooling} & D \details{adversarial example} & I \details{ImageNet, VOC2012} & N \details{VGG, ResNet, DenseNet} & L \details{Grad-CAM} \\
        \citet{heo2019fooling} & M \details{model manipulation} & I \details{ImageNet} & N \details{VGG, ResNet, DenseNet} & L \details{SG, Grad-CAM, LRP} \\
        \citet{dombrowski2019explanations} & D \details{adversarial example} & I \details{ImageNet, CIFAR-10} & N \details{VGG, ResNet, DenseNet, ..} & L \details{SG, GI, IG, LRP, ..} \\
        \citet{dimanov2020you} & M \details{model manipulation} & T \details{Credit, COMPAS, Adult, ..} & N \details{MLP} & L \details{SG, GI, IG, SHAP, ..} \\
        \citet{slack2020fooling} & M \details{adversarial model} & T \details{Credit, COMPAS, Crime} & B \details{rule set} & L \details{SHAP, LIME} \\
        \citet{lakkaraju2020how} & E \details{trust manipulation} & T \details{Bail} & B \details{rule set} & G \details{MUSE} \\
        \citet{anders2020fairwashing} & M \details{model manipulation} & T \details{credit,} I \details{MNIST, CIFAR10, ..} & N \details{LR, CNN, VGG} & L \details{SG, GI, IG, LRP} \\
        \citet{kuppa2020blackbox} & D \details{adversarial example} & T \details{PDF, Android, UGR16} & N \details{MLP, GAN} & L \details{SG, GI, IG, LRP, ..} \\
        \citet{zhang2020interpretable} & D \details{adversarial example} & I \details{ImageNet} & N \details{ResNet, DenseNet} & L \details{SG, CAM, RTS, ..} \\
        \citet{merrer2020remote} & M \details{adversarial model} & T \details{Credit} & B \details{DT, MLP} & L \details{custom} \\
        \citet{shokri2021privacy} & -- \details{membership inference} & T \details{Adult, ..} I \details{CIFAR-10, ..} & B \details{MLP, CNN} & L \details{IG, LRP, LIME, ..} \\
        \citet{sinha2021perturbing} & D \details{adversarial example} & Lg \details{IMDB, SST, AG News} & B \details{DistilBERT, RoBERTa} & L \details{IG, LIME} \\
        \citet{zhang2021data} & M \& D \details{data poisoning} & T \details{Fracture,} I \details{Dogs} & N \details{MLP, ResNet} & L \details{SG, CAM} \\
        \citet{slack2021counterfactual} & M \details{model manipulation} & T \details{Credit, Crime} & N \details{MLP} & L \details{counterfactual} \\
        \citet{baniecki2022manipulating} & D \details{data poisoning} & T \details{Heart, Apartments} & B \details{XGBoost} & G \details{SHAP,} L \details{SHAP}\\
        \citet{brown2022making} & D \details{data poisoning} & I \details{ImageNet, CUB} & N \details{InceptionNet, ResNet, ViT, ..} & G \details{TCAV, FFV} \\
        \citet{baniecki2022fooling} & D \details{data poisoning} & T \details{Heart, Friedman} & B \details{MLP, RF, GBDT, SVM, ..} & G \details{PDP} \\
        \citet{pawelczyk2023privacy} & -- \details{membership inference} & T \details{Adult, Hospital} & B \details{LR, NN} & L \details{counterfactual} \\
        \citet{laberge2023fooling} & D \details{data poisoning} & T \details{COMPAS, Adult, Bank, Crime} & B \details{MLP, RF, XGBoost} & G \details{SHAP} \\
        \citet{noppel2023disguising} & M \& D \details{backdoor attack} & I \details{CIFAR-10, GTSRB} & N \details{ResNet} & L \details{SG, Relevance-CAM, ..} \\
        \citet{tamam2023foiling} &  D \details{adversarial example} & I \details{ImageNet, CIFAR-10} & B \details{VGG, MobileNet, ..} & L \details{SG, GI, DeepLIFT, ..} \\
        \citet{huang2023safari} & D \details{adversarial example} & I \details{MNIST, CIFAR-10, CelebA} & B \details{CNN, ResNet, MobileNet} & L \details{GI, IG, LRP, LIME, ..} \\
        \bottomrule
    \end{tabular}
\end{table*}

\paragraph{Adversarial model} In a black-box setting, \citet{slack2020fooling} manipulate LIME and SHAP explanations for tabular data by exploiting their reliance on perturbing input data for estimation. The proposed attack substitutes a biased black-box with a model surrogate to effectively hide bias, e.g. from auditors. In detail, an out-of-distribution detector is trained to divide input data such that the black-box'es predictions in-distribution remain biased, but its behavior on the perturbed data is controlled, which makes the explanations look fair:
\begin{equation}
    f \rightarrow f'
        \Longrightarrow 
    \left\{\begin{array}{@{}l@{}}
        \exists_{\bx \in \bX}\;g(f,\bx) \neq g(f',\bx) \\  
        \forall_{\bx \in \bX}\;f(\bx) \approx f'(\bx) \\
    \end{array}\right..
\end{equation}
Figure~\ref{fig:fooling-shap} shows exemplary explanations resulting from such an adversarial model attack. The most important feature in model $f$ becomes less important in model $f'$ as judged by the attacked explanations. \citet{merrer2020remote} consider a similar adversarial scenario where the user queries a black-box service, e.g. a company, to obtain an explanation of a decision. A malicious service provider hides biased predictions by returning fairly looking explanations of a surrogate model. In theory, allowing to query multiple explanations cannot prevent a remote service from hiding the true reasons that lead to its decision. However, experiments show that an impractically large number of user queries is required to detect such explanation manipulation.

\paragraph{Trust manipulation} While the majority of adversarial attacks are on local methods for interpreting individual predictions, other attacks specifically target global methods explaining the overall model's reasoning \citep{lakkaraju2020how,baniecki2022manipulating,brown2022making,baniecki2022fooling,laberge2023fooling}. Instead of changing model or data, \citet{lakkaraju2020how} introduce misleading rule-based explanations that approximate a model based on the MUSE framework \citep{lakkaraju2019faithful}. Results of a user study show that various high-fidelity explanations faithful to the black-box considerably affect human judgement: 
\begin{equation}
    g \rightarrow g'
        \Longrightarrow 
    \left\{\begin{array}{@{}l@{}}
        g(f,\bX) \neq g'(f,\bX) \\  
        \forall_{\bx \in \bX}\;g'(\bx) \approx f(\bx) \\
    \end{array}\right..
\end{equation}
Constructing differing faithful explanations is possible due to the correlation between features. User trust in the model decreases when sensitive information appears in an explanation, e.g. gender and race, and it increases when domain-specific features appear as important~\citep{lakkaraju2020how}.

\begin{figure}[t]
    \centering
    \includegraphics[width=0.59\textwidth]{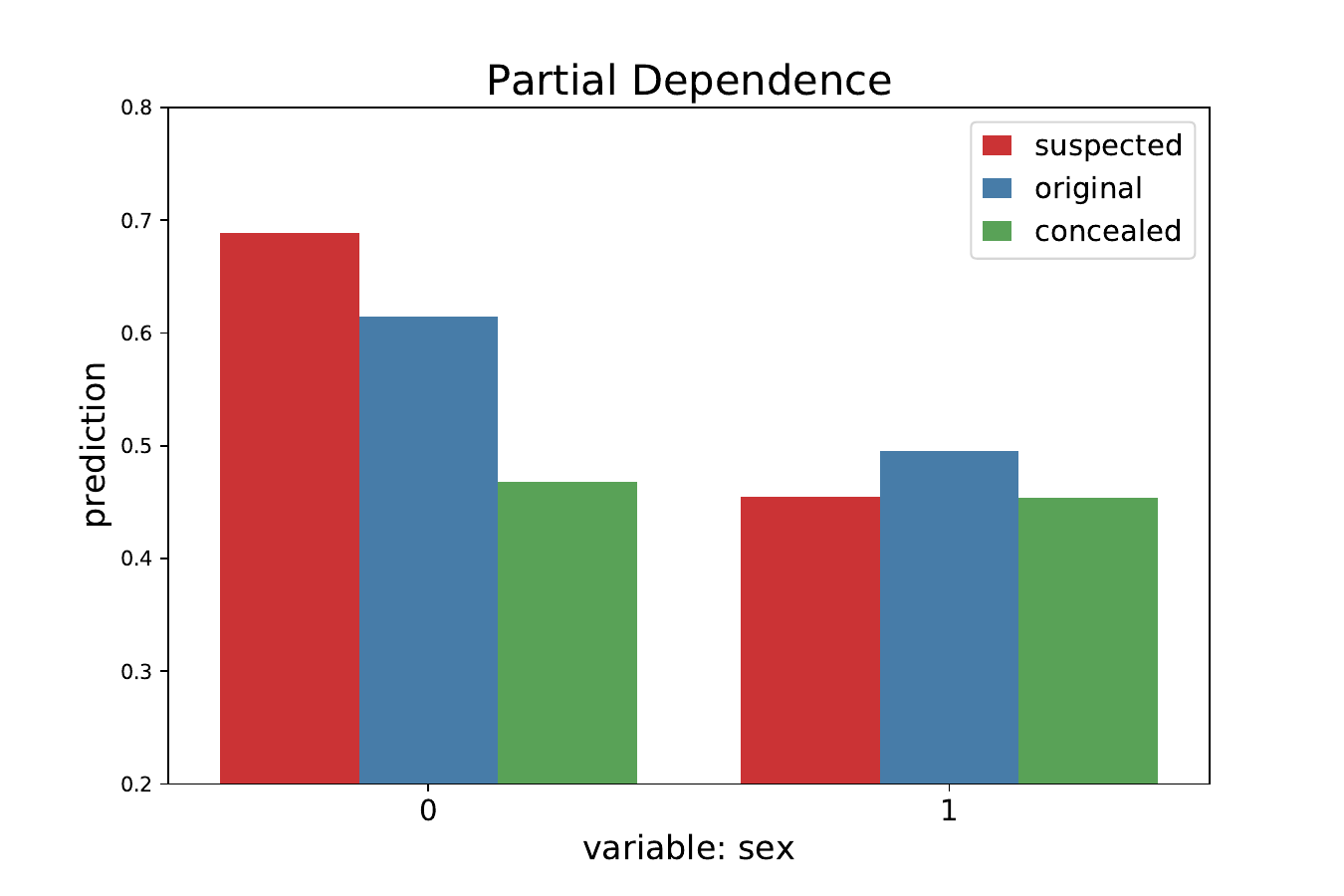}
    \includegraphics[width=0.40\textwidth]{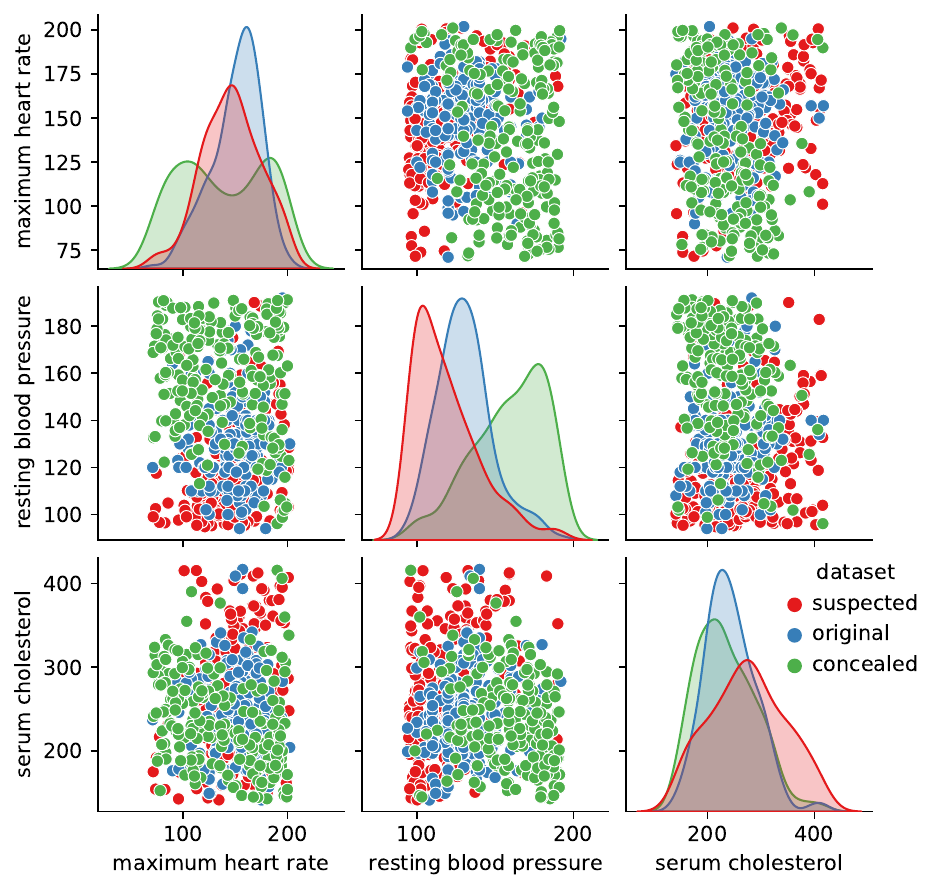}
    \caption{Data poisoning attack on feature effect explanations of a machine learning model predicting a heart attack risk (class~0)~\citep[adapted from][]{baniecki2022fooling}. \emph{Left:} An attacker manipulates explanations to present either a suspected or concealed contribution of feature ``sex'' to the predicted outcome. \emph{Right:} Distribution of the 3 poisoned features from the data, in which ``sex'' and the remaining 9 features attributing to the explanation remain unchanged. }\label{fig:fooling-pd}
\end{figure}

\paragraph{Data poisoning} \citet{baniecki2022manipulating} and \citet{baniecki2022fooling} introduce genetic-based algorithms to manipulate SHAP and PDP explanations respectively. The proposed poisoning attack iteratively changes data used in the process of estimating global explanations, and thus can be exploited by an adversary to provide false evidence of feature importance and effects: 
\begin{equation}
    \bX \rightarrow \bX'
        \Longrightarrow
    g(f,\bX) \neq g(f,\bX'). 
\end{equation}
Figure~\ref{fig:fooling-pd} shows exemplary explanations resulting from such a data poisoning attack. The sensitive feature in model $f$ becomes less important in explanation computed on dataset $X'$ as judged by the attacked explanations.
\citet{laberge2023fooling} consider a similar adversarial scenario and attack global SHAP with biased sampling of the data points used to approximate explanations \citep[an algorithm introduced in][]{fukuchi2020faking}. It is done stealthily aiming to minimize the difference in data distributions. Experiments show an improvement in manipulating SHAP over previous work of \citet{baniecki2022manipulating}, which further underlines SHAP' vulnerability~\citep{slack2020fooling}. \citet{tamam2023foiling} and \citet{huang2023safari} use genetic-based algorithms as a black-box attack to craft adversarial examples for local post-hoc explanations of image classification.

\paragraph{Membership inference} We further acknowledge that instead of attacking explanations, they might be exploited to breach privacy in machine learning applications. \citet{shokri2021privacy} introduce membership inference attacks that use information from feature attributions to determine whether a data point was present in the training dataset. \citet{pawelczyk2023privacy} propose membership inference attacks using counterfactual explanations instead. While many contributions consider the detectability of the attack \citep{subramanya2019fooling,kuppa2020blackbox,zhang2020interpretable}, and some propose ways of mitigating the attacks' effects via robustifying mechanisms \citep{dombrowski2019explanations,anders2020fairwashing,zhang2020interpretable}, we systemize contributions that mostly focus on defending in Section~\ref{sec:defense}.

\paragraph{Insights} The majority of proposed attacks assume prior knowledge that the explained model is a neural network, e.g. to utilize gradient descent in constructing adversarial examples or changing model parameters, as opposed to black-box approaches that could work with various model algorithms. Attacking local explanations may impact the model's behaviour globally~\citep{heo2019fooling,dimanov2020you,anders2020fairwashing,noppel2023disguising}, and vice versa, fooling global explanations may manipulate local explanations in the process~\citep{lakkaraju2020how,baniecki2022manipulating,laberge2023fooling}. Both of these interactions could improve detectability in practice. To this date, there are relatively sparse studies concerning adversarial attacks on concept-based explanations, e.g.~\citet{brown2022making} attack TCAV~\citep{kim2018interpretability} and FFV~\citep{goh2021multimodal}, counterfactual explanations, e.g. \citet{slack2021counterfactual} attack counterfactuals for neural networks~\citep{guidotti2022counterfactual}, and overall explanations for language models, e.g. \citet{sinha2021perturbing} attack IG and LIME. Research on attacking explanations for image classification relies on a few popular datasets, e.g. ImageNet~\citep{deng2009imagenet} reoccurs in 8 out of 11 studies, while a larger variety of tabular scenarios is tested.

\section{Defense against the attacks on explanations}\label{sec:defense}

Whenever a new attack algorithm is introduced in adversarial machine learning, various ways to address the explanation's limitations and fix its insecurities are proposed. \citet{chen2019robust} is one of the first attempts to defend from adversarial examples introduced by \citet{ghorbani2019interpretation} via regularizing a neural network. The proposed robust attribution regularization forces integrated gradients explanations to remain unchanged under perturbation attacks. \citet{rieger2020simple} propose an alternative defense strategy against such adversarial examples \citep{ghorbani2019interpretation,dombrowski2019explanations} by aggregating multiple explanations created with various algorithms. Specifically, feature importance scores estimated with different heuristics are averaged to obtain a more robust explanation. As the attack targets only a single explanation method, their ensemble remains close to the original explanation (shown in Figure~\ref{fig:adversarial-example}).
 
Table~\ref{tab:defenses} lists defenses against the attacks on explanations, where for each, we record the datasets, models and explanation algorithms mentioned in experiments. Excluded from it are works that improve explanation robustness without directly relating to the potential adversarial attack scenario \citep[e.g. see][and references given there]{yeh2019infidelity,zhou2021slime,zhao2021baylime,slack2021reliable,meyer2023minimizing}.

\begin{table*}[!ht]
    \setlength{\tabcolsep}{2.5pt}
    \centering
    \caption{Summary of works introducing defense mechanisms against the attacks on explanations of machine learning models. Each work is connected with \emph{up to two} attacks that are mentioned to be potentially addressed by it. Experiments consider data modalities including image~(I), tabular~(T), language~(Lg); and model algorithms specifically neural networks~(N), or black-box models~(B) including neural networks. Explanation methods are either local (L) or global~(G). \ref{app:abbr} lists other abbreviations.}
    \label{tab:defenses}
    \vspace{1em}
    \begin{tabular}{@{\hspace{1pt}}LLLLL@{\hspace{1pt}}}
        \toprule
         \textbf{Defense} & \textbf{Attack(s)} & \textbf{Modality \details{dataset}} & \textbf{Model \details{algorithm}} & \textbf{Explanation \details{method}} \\
        \midrule
        \citet{woods2019adversarial} & -- & I \details{ImageNet, COCO, ..} & N \details{ResNet} & L \details{Grad-CAM} \\
            \greyrule
        \citet{chen2019robust} & \citet{ghorbani2019interpretation} & I \details{FashionMNIST, Flower, ..} & N \details{CNN, ResNet} & L \details{IG} \\
            \greyrule
        \citet{rieger2020simple} & \makedoublecell{\citet{ghorbani2019interpretation}}{\citet{dombrowski2019explanations}} & I \details{ImageNet} & N \details{VGG} & L \details{SG, IG, LRP, GBP} \\
            \greyrule
        \citet{boopathy2020proper} & \makedoublecell{\citet{ghorbani2019interpretation}}{\citet{dombrowski2019explanations}} & I \details{MNIST, CIFAR-10, ..} & N \details{CNN, ResNet} & L \details{IG, CAM, Grad-CAM} \\
            \greyrule
        \citet{lakkaraju2020robust} & \makedoublecell{\citet{ghorbani2019interpretation}}{\citet{lakkaraju2020how}} & T \details{Bail, Academic, Health} & B \details{MLP, RF, GBDT, ..} & G \details{MUSE, LIME, SHAP} \\
            \greyrule
        \citet{wang2020smoothed} & \makedoublecell{\citet{ghorbani2019interpretation}}{\citet{dombrowski2019explanations}} & I \details{CIFAR-10, ImageNet, Flower} & N \details{ResNet} & L \details{SG, IG, SmoothGrad} \\
            \greyrule
        \citet{lamalfa2021guaranteed} & -- & Lg \details{IMDB, SST, Twitter} & N \details{MLP, CNN} & L \details{Anchors} \\
            \greyrule
        \citet{ghalebikesabi2021locality} & \citet{slack2020fooling} & \makedoublecell{T \details{COMPAS, Adult, Bike, ..}}{I \details{MNIST}} & B \details{XGBoost, CNN} & L \details{SHAP, GradSHAP} \\
            \greyrule
        \citet{dombrowski2022towards} & \citet{dombrowski2019explanations} & I \details{CIFAR-10, ImageNet} & N \details{CNN, VGG, ResNet} & L \details{SG, GI, IG, LRP, ..} \\
            \greyrule
        \citet{schneider2022deceptive} & -- & Lg \details{IMDB, WoS} & B \details{CNN} & L \details{Grad-CAM} \\
            \greyrule
        \citet{tang2022defense} & \makedoublecell{\citet{ghorbani2019interpretation}}{\citet{dombrowski2019explanations}} & I \details{MNIST, FashionMNIST} & N \details{CNN} & L \details{SG} \\
            \greyrule
        \citet{shrotri2022constraint} & \citet{slack2020fooling} & T \details{Credit, COMPAS, Crime, ..} & B \details{RF} & L \details{LIME} \\
            \greyrule
        \citet{gan2022is} & \makedoublecell{\citet{dombrowski2019explanations}}{\citet{zhang2020interpretable}} & \makedoublecell{I \details{ImageNet, CityScapes}}{Lg \details{IMDB, Toxic}} & NN \details{ResNet, LSTM, ..} & L \details{CAM, LRP, LIME, ..} \\
            \greyrule
        \citet{vres2022preventing} & \citet{slack2020fooling} & T \details{Credit, COMPAS, Crime} & B \details{MLP, RF, SVM, ..} & L \details{LIME, SHAP, IME} \\
            \greyrule
        \citet{liu2022certifiably} & \citet{ghorbani2019interpretation} & I \details{VOC2007} & N \details{VGG} & L \details{SG} \\
            \greyrule
        \citet{carmichael2023unfooling} & \citet{slack2020fooling} & T \details{Credit, COMPAS, Crime} & B \details{rule set} & L \details{SHAP, LIME} \\
            \greyrule
        \citet{joo2023towards} & \makedoublecell{\citet{ghorbani2019interpretation}}{\citet{dombrowski2019explanations}} & I \details{CIFAR-10, ImageNet} & N \details{ResNet, LeNet} & L \details{SG, GI, LRP, GBP} \\
            \greyrule
        \citet{virgolin2023robustness} & \citet{slack2021counterfactual} & T \details{Credit, COMPAS, Adult, ..} & B \details{MLP, RF} & L \details{counterfactual}  \\
            \greyrule
        \citet{wicker2023robust} & \makedoublecell{\citet{heo2019fooling}}{\citet{dombrowski2019explanations}} & \makedoublecell{T \details{Credit, Adult,}}{I \details{MNIST, MedMNIST}} & N \details{MLP, CNN} & L \details{GI, DeepLIFT, SHAP} \\
            \greyrule
        \citet{pawelczyk2023probabilistically} & \citet{slack2021counterfactual} & T \details{Credit, COMPAS, Adult} & N \details{LR, MLP} & L \details{counterfactual} \\
            \greyrule
        \citet{blesch2023unfooling} & \citet{slack2020fooling} & T \details{Credit} & B \details{rule set} &  G \details{SHAP,} L \details{SHAP} \\
        \bottomrule
    \end{tabular}
\end{table*}

\paragraph{Pairing defenses and attacks} We link each defense with an attack, but omit to list all attacks potentially addressed by the defense for brevity. The three missing links are worth clarifying here. \citet{woods2019adversarial} is an early work that introduces adversarial explanations, which have improved robustness against adversarial examples targeting model predictions. Similarly, \citet{lamalfa2021guaranteed} proposes to improve explanations of language models against adversarial perturbations. They consider explanations in the form of a minimal subset of words (tokens in general) that are sufficient to imply the model's prediction, e.g. a positive sentiment of an input sentence. Explanation robustness is defined as an invariance with respect to bounded perturbations to the remaining words. Two algorithms are used to construct explanations satisfying such constraints: minimum hitting set~\citep{ignatiev2019abduction} and minimum satisfying assignment~\citep{dillig2012minimum}. Unlike most of the contributions that focus on algorithms, \citet{schneider2022deceptive} conduct a user study with artificially manipulated explanations to evaluate if humans can be deceived by the potential attack \citep[related to][]{lakkaraju2020how,poursabzi2021manipulating}. The task is to decide whether the model’s prediction of a movie review sentiment is correct. Results of 3500 answers from 140 crowdsourced participants show that many humans can be deceived by explanation manipulation. \citet{virgolin2023robustness} and \citet{pawelczyk2023probabilistically} introduce mechanisms to improve the robustness of counterfactual explanations against adversarial perturbations (we link the latter with an otherwise unreferenced attack of \citet{slack2021counterfactual}). The general idea is to find counterfactual input examples that are robust to adversarial perturbations. It is possible to penalize the loss function used to optimize counterfactual explanations for modifying values of features, which are most vulnerable to perturbations~\citep{virgolin2023robustness}. \citet{pawelczyk2023probabilistically} propose a novel recourse invalidation rate that measures an expected distance between the prediction for a counterfactual example and inputs in its neighbourhood. Optimizing it naturally leads to counterfactual explanations robust to perturbations.

\paragraph{Defense of explanations specific to neural networks} \citet{boopathy2020proper} extend the regularization training method of \citet{chen2019robust} to use an $l_1$-norm 2-class interpretation discrepancy measure. Experiments show an improvement in effectiveness and computation cost when defending explanations over previous work~\citep[including][]{chen2019robust}. Moreover, the proposed model regularization that improves explanation robustness also improves robustness to adversarial examples that target model predictions. \citet{wang2020smoothed} introduce a smooth surface regularization procedure to force robust attributions by minimizing the difference between explanations for nearby points. Experiments show a trade-off between regularization performance and computation cost \citep[also with respect to][]{chen2019robust}. Notably, models with smoothed geometry become less susceptible to transfer attacks, i.e. where an adversary targeting one explanation method fools other gradient-based explanations as well. \citet{dombrowski2022towards} and \citet{tang2022defense} further compare and extend the in-training techniques to regularize neural networks towards improving explanation robustness~\citep[also mentioned in][]{dombrowski2019explanations}. \citet{dombrowski2022towards} use the approximated norm of the Hessian as a regularization term during training to bound the $l_2$-distance between the gradients of the original and perturbed inputs, which benefits gradient-based explanations. \citet{joo2023towards} improve this approach by introducing a cosine robust criterion to measure the $cosine$-distance instead. Minimizing the $cosine$-distance between the gradients of inputs sampled near training inputs promotes locally aligned explanations. As shown in experiments comparing the two distance measures, $cosine$-distance effectively solves the issue of $l2$-distance criterion being not normalization invariant, which is important as normalized attribution values are often used to analyse explanations in practice. \citet{gan2022is} use hypothesis testing to quantify the uncertainty of feature attributions and increase their stability in adversarial scenarios. The proposed MeTFA framework can be applied on top of various explanation methods, relating closely to SmoothGrad~\citep{smilkov2017smoothgrad}. Experiments across image and text classification tasks show the superiority of MeTFA over SmoothGrad, as well as its utility in defending from adversarial examples~\citep{zhang2020interpretable}. Most recent work in this line of research introduces certifiably robust explanations of neural networks~\citep{liu2022certifiably,wicker2023robust}, which reassure that no adversarial explanation exists for a given set of input or model weights.

\paragraph{Defense of model-agnostic explanations} In parallel to defending adversarial attacks on gradient-based explanations of neural networks, several works address the possibility of fooling model-agnostic LIME and SHAP~\citep{slack2020fooling}. \citet{ghalebikesabi2021locality} modifies the SHAP estimator by sampling data from a local neighbourhood distribution instead of the marginal or conditional global reference distribution. Experiments show that such constructed on-manifold explainability improves explanations' robustness, i.e. SHAP defends from the attack. \citet{shrotri2022constraint} modifies the LIME estimator to take into account user-specified constraints on the input space that restrict the allowed data perturbations. Alike, experiments show that constrained explanations are less susceptible to out-of-distribution attacks. Moreover, analysing differences between the original and constrained explanations allows for detecting an adversarially discriminative classifier. In contrast, \citet{vres2022preventing} introduce focused sampling with various data generators to improve the adversarial robustness of both LIME and SHAP. Instead of directly improving perturbation-based explanation methods, \citet{carmichael2023unfooling} propose to ``unfool'' explanations with conditional anomaly detection. An algorithm based on k-nearest neighbours scores the abnormality of inputs conditioned on their classification labels. Comparing the empirical distribution function of scores between the original and potentially adversarially perturbed inputs given a user-defined threshold proves to be effective for attack detection. Removing abnormal inputs from the perturbed set defends an explanation against fooling (see an example in Figure~\ref{fig:fooling-shap}). \citet{blesch2023unfooling} propose to impute out-of-coalition features with model-X knockoffs in calculating SHAP values, which avoids extrapolation and effectively defends against the adversarial attack~\citep{slack2020fooling}. It extends this methodology to improve the robustness of global feature importance based on SHAP~\citep{covert2020understanding}. Most recently, \citet{lin2023robustness} derive theoretical bounds for the robustness of feature attribution explanations to input and model perturbations.

\paragraph{Insights} Comparing Tables~\ref{tab:attacks}~\&~\ref{tab:defenses} highlights existing insecurities in XAI methods; namely, not clearly addressed are backdoor attacks~\citep{viering2019how,noppel2023disguising}, data poisoning attacks~\citep{zhang2021data,baniecki2022fooling}, attacks specific to language~\citep{sinha2021perturbing} and concept-based explainability~\citep{brown2022making}. To this date, there are sparse studies concerning defenses against attacks on global explanations, e.g. \citet{lakkaraju2020robust} robustify model-agnostic global explanations against a general class of distribution shifts related to adversarial perturbations~\citep{ghorbani2019interpretation}.

\section{Adversarial attacks on fairness metrics}\label{sec:fairness}

Closely related to adversarial attacks on explanations are attacks on machine learning fairness metrics, e.g. predictive equality~\citep{corbett2017algorithmic} and (statistical) demographic parity~\citep[refer to][for an introduction to machine learning fairness]{mehrabi2021survey}. Intuitively, algorithms targeting model predictions and accuracy can be applied to manipulate other functions of the model output as well. Table~\ref{tab:fairness} lists a representative set of adversarial attacks on fairness metrics, with the corresponding strategy of changing data~\citep{fukuchi2020faking}, the model~\citep{aivodji2019fairwashing,aivodji2021characterizing}, or jointly changing data and the model~\citep{solans2020poisoning,mehrabi2021exacerbating,hussain2022adversarial}. 

\begin{figure}[t]
    \centering
    \includegraphics[width=0.99\linewidth]{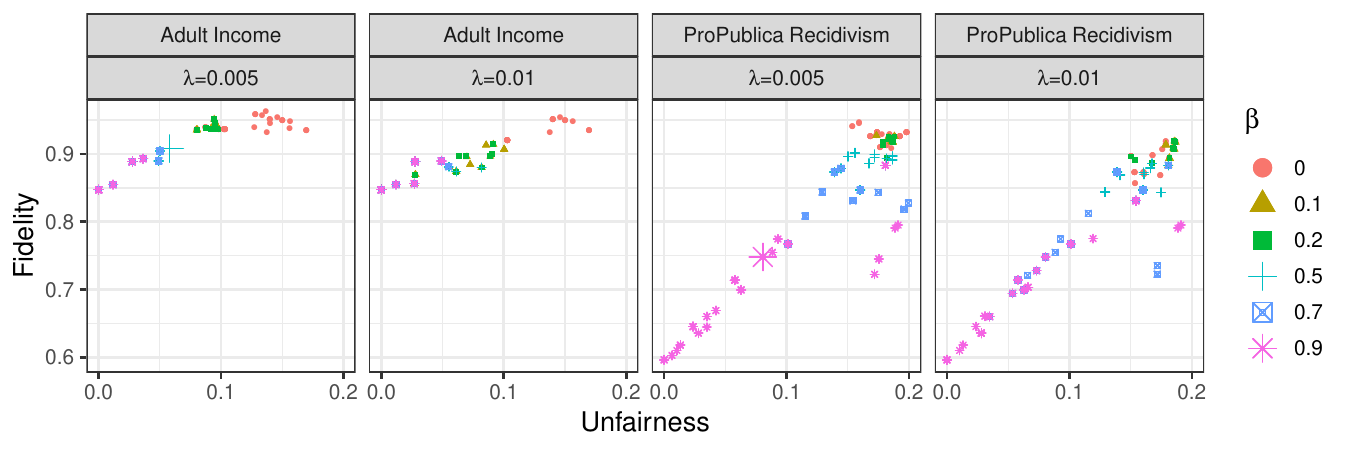}
    \caption{Adversarial model attack on fairness measurements of machine learning models predicting income and recidivism \citep[adapted from][]{aivodji2019fairwashing}. Multiple surrogate rule list models (points) are fitted to faithfully approximate the unfair black-box predictive function. Parameter $\lambda$ controls the penalty for rule length (complexity) and $\beta$ controls a trade-off between fidelity and unfairness measured with statistical parity.}
    \label{fig:fairwashing}
\end{figure}

\paragraph{Adversarial model} \citet{aivodji2019fairwashing,aivodji2021characterizing} introduce fairwashing attacks changing the model, i.e. an adversary approximates an unfair black-box model with a faithful adversarial model appearing as fair:
\begin{equation}
    f \rightarrow f'
        \Longrightarrow 
    \left\{\begin{array}{@{}l@{}}
        g(f,\bX) \neq g(f',\bX) \\  
        \forall_{\bx \in \bX}\;f(\bx) \approx f'(\bx) \\
    \end{array}\right..
\end{equation}
Experiments analyse the fidelity-unfairness trade-off and the effect of fairwashing on impacting feature effect explanations (see Figure~\ref{fig:fairwashing}). 

\paragraph{Data poisoning} \citet{fukuchi2020faking} introduce a stealthily biased sampling procedure to adversarially craft an unbiased dataset used to estimate fairness metrics. It is formally defined as a Wasserstein distance minimization problem and solved with an efficient algorithm for a minimum-cost flow problem in practice. Experiments focus on quantifying the trade-off between lowering the perceived model bias and detecting adversarial sampling. Contrary, \citet{solans2020poisoning} introduce a data poisoning attack to increase bias as measured with fairness metrics via adding data points to the training dataset so that the model discriminates against a certain group of individuals: 
\begin{equation}
    \bX \rightarrow \bX'
        \Rightarrow 
    f_{\theta} \rightarrow f_{\theta'}
        \Longrightarrow
    g(f_{\theta},\bX) \neq g(f_{\theta'},\bX).
\end{equation}
Although the approach relies on the differentiation of neural networks for optimization, experiments show the transferability of data poisoning to other model algorithms in a so-called black-box setting where the attacker has no access to the model. \citet{mehrabi2021exacerbating} propose alternative data poisoning attacks on fairness: a gradient-based influence attack~\citep{koh2021stronger}, which was previously proposed for attacking predictive performance, and a novel anchoring attack applicable to any black-box predictive function. The intuition behind an anchoring attack is to manipulate the model's decision boundary by generating points that have the same demographic group but opposite labels. Experiments show an improvement in manipulating fairness over previous work of \citet{solans2020poisoning}, also in the black-box setting. \citet{hussain2022adversarial} extend data poisoning attacks on fairness to the task of node classification with graph convolutional networks. The attacker's strategy is to add adversarial edges into a graph (training dataset) to increase the model's bias. Results confirm that graph neural networks are vulnerable to degrading fairness metric values without a significant drop in predictive performance.

\begin{table*}[t]
    \centering
    \caption{Related work concerning adversarial attacks on fairness metrics for measuring bias of machine learning models. Most of the attacks change at least one of the data~(D) or model~(M). Experiments consider data modalities including image~(I), tabular~(T), graph~(Gr); and model algorithms specifically neural networks~(N), or black-box models~(B) including neural networks. \ref{app:abbr} lists other abbreviations.}
    \label{tab:fairness}
    \vspace{1em}
    \begin{tabular}{@{\hspace{1pt}}LLLLL@{\hspace{1pt}}}
        \toprule
        \textbf{Attack} & \textbf{Changes \details{strategy}} & \textbf{Modality \details{dataset}} & \textbf{Model \details{algorithm}} & \textbf{Fairness \details{metric}}\\
        \midrule
        \citet{aivodji2019fairwashing} & M \details{adversarial model} & T \details{COMPAS, Adult} & B \details{RF} & \details{SP} \\
        \citet{fukuchi2020faking} & D \details{data poisoning} & T \details{COMPAS, Adult} & B \details{RF, LR} & \details{SP}\\
        \citet{solans2020poisoning} & D \& M \details{data poisoning} & T \details{COMPAS} & B \details{LR, RF, SVM, DT, naïve Bayes} & \details{SP, EOdds} \\
        \citet{mehrabi2021exacerbating} & D \& M \details{data poisoning} & T \details{Credit, COMPAS, Drug} & B \details{MLP} & \details{SP, EOdds} \\
        \citet{nanda2021fairness} & D, M \details{adversarial example} & I \details{Adience, UTKFace, ..} & N \details{VGG, ResNet, DenseNet, ..} & \details{robustness bias} \\
        \citet{aivodji2021characterizing} & M \details{adversarial model} & T \details{Credit, COMPAS, Adult, ..} & B \details{MLP, RF, AdaBoost, XGBoost} & \details{SP, PE, EOdds, EOpp}\\
        \citet{hussain2022adversarial} & D \& M \details{data poisoning} & Gr \details{Pokec, DBLP} & N \details{GCN} & \details{SP, EOdds, EOpp} \\
        \citet{ferry2023exploiting} & -- \details{data reconstruction} & T \details{ACSIncome, ACSPublicCoverage} & B \details{DT+\texttt{fairlearn}} & \details{SP, PE, EOdds, EOpp}\\
        \bottomrule
    \end{tabular}
\end{table*}

\paragraph{Adversarial example} Further related is work concerning attacking fairness in imaging, where \citet{nanda2021fairness} introduce the notion of \emph{robustness bias}. Robustness bias concerns scenarios where groups of inputs are inequally susceptible to adversarial attacks, e.g. adversarial examples. 

\paragraph{Data reconstruction} Instead of attacking fairness metrics, \citet{ferry2023exploiting} consider a different adversarial scenario where the goal of an attacker is to retrieve a piece of information about a sensitive attribute based on metric values (dataset extraction attack). Related are membership inference attacks on privacy using explanations~\citep{shokri2021privacy,pawelczyk2023privacy}.

\paragraph{Insights} Research on attacking fairness metrics focuses on the notion of \emph{group fairness}, i.e. treating different groups of inputs equally, and omits subgroup or \emph{individual fairness}, i.e. predicting similarly for similar individuals~\citep[see the distinction in][table 1]{mehrabi2021survey}.

\section{AdvXAI: Opportunities, challenges and future research directions} \label{sec:discussion}

As the popularity of explainable AI grew, explanation methods left the lab and found their way into potentially hostile habitats. In order to defend their position as guardians of the trust and security of AI systems, explanations must also demonstrate their resilience to adversarial action. Below, we outline the potential short-term and long-term research directions in the AdvXAI field.

\paragraph{Attacks} Currently most exploited by the attacks are the first-introduced and most popular XAI methods, e.g. SHAP and Grad-CAM. Future work on adversarial attacks may consider targeting the more recent enhancements that aim to overcome their limitations, e.g. SHAPR that takes into account feature dependence in tabular data~\citep{aas2021explaining} or Shap-CAM for improved explanations of convolutional neural networks~\citep{zheng2022shap}. Alike model-specific explanations of neural networks, worth assessing is the vulnerability of explanation methods specific to tree-based models, e.g. TreeSHAP~\citep{lundberg2020from}, but also attacks on model-agnostic explanations assuming prior knowledge that the model is an ensemble of decision trees. Beyond post-hoc explainability, adversarial attacks could target vulnerabilities of the interpretable by-design deep learning models like ProtoPNet~\citep{chen2019this} and its extensions~\citep[e.g. see][and its related work]{rymarczyk2022interpretable}. Finally, there are adversarial attacks on model predictions that actively aim to bypass through a particular defense mechanism \citep[e.g. see][table 4]{machado2021adversarial} and such a threat of circumventing the defense in XAI is currently unexplored. 

\paragraph{Defenses} One goal of this survey is to reiterate the apparent insecurities in XAI, i.e. the unaddressed attacks on explanation methods~\citep{viering2019how,zhang2021data,brown2022making,baniecki2022fooling,noppel2023disguising,laberge2023fooling}. Beyond defense, improving the robustness and stability of algorithms for estimating explanations becomes crucial, e.g. gradient-based~\citep{yeh2019infidelity} and perturbation-based~\citep{zhou2021slime,zhao2021baylime,slack2021reliable} methods. \citet{meyer2023minimizing} provide theoretical and empirical characterizations of the factors influencing explanation stability under data shifts. We also underline that the possibility of manipulating fairness metrics has detrimental consequences when applied in audit and law enforcement, and therefore developing metrics robust against the attacks is desirable. Note that although a particular XAI method is attacked or defended, in fact, the evidence of model predictions is in question here.

\paragraph{AdvXAI beyond classical models towards transformers}
Nowadays, the transformer architecture is at the frontier of machine learning research and applications of deep learning in practice~\citep{komorowski2023towards}. Thus, the adversarial robustness of explanations of large models for various modalities like ViT~\citep{dosovitskiy2021transformers} and TabPFN~\citep{hollmann2023tabpfn} deserves special attention. For example, \citet{ali2022xai} extend LRP explanations to transformers, which might propagate the explanations' vulnerability to adversarial attacks as shown in previous work~\citep{heo2019fooling,anders2020fairwashing}. We acknowledge that the recently proposed transformer-based foundation models, e.g. SAM~\citep{alex2023segment}, more and more frequently include benchmarks specific to evaluating responsibility, e.g. whether segmenting people from images is unbiased with respect to their perceived gender presentation, age group or skin tone~\citep{schumann2021step}. The possibility of attacking such fairness measurements by biased sampling becomes a trust issue~\citep{fukuchi2020faking}. 

\paragraph{AdvXAI beyond the image and tabular data modalities}
A majority of contributions surveyed here, so as XAI, concern machine learning predictive models trained on imaging and tabular datasets. Further work is required to evaluate which and how severe are adversarial attacks concerning other data modalities like language~\citep{lamalfa2021guaranteed,schneider2022deceptive}, graphs~\citep{hussain2022adversarial}, time series, multimodal systems, and explanations of reinforcement learning agents~\citep{madumal2020explainable,olson2021counterfactual}. \citet{huai2020malicious} are the first to introduce a model poisoning attack against the explanations of deep reinforcement learning agents, with currently no consecutive directly related work.

\paragraph{Software, datasets and benchmarks} To facilitate sustainable research in AdvXAI, one could develop open-source software implementing the reviewed attacks and defenses, which promotes reproducibility and a unified comparison between methods going forward. To this end, software contributions concerning responsible and secure machine learning~\citep{baniecki2021dalex,pintor2022secml,hedstrom2023quantus,weerts2023fairlearn} lack the implementation of adversarial attacks on explanations and fairness metrics. Moreover, contributing new real-world datasets and benchmarks specifically aiming at evaluating attacks and defenses in XAI would be valuable~\citep[alike][]{schumann2021step,arras2022clevrxai,agarwal2022openxai}. An implementation of such a security benchmark could resemble the Safety Gym framework for testing safe exploration in reinforcement learning \citep{ray2019benchmarking}. 
Alongside, competitions have proven to accelerate research in particular domains of machine learning, e.g. the FICO competition for XAI~\citep{fico2018explainable}.

\paragraph{Good practices} Historically, machine learning models were evaluated in train and test settings of data distribution. Adversarial methods broadened the scope of evaluation into adversarial settings, e.g. out-of-distribution data~\citep{meyer2023minimizing}. While there is a need for discussion on whether explanations should be evaluated on train or test data, new adversarial settings described in this survey need to be also taken into account. Another good practice would be to catalogue the collective history of harms or near harms realized in the real world by the deployment of XAI methods, similar to the AI Incident Database~\citep{partnership2023database}. A step further, various certifications of AI systems appear in the context of their deployment in real-world applications~\citep{cihon2021certification}, which could be applicable to XAI, apart from algorithmic constraints-driven certificates for explanation methods \citep{gu2020certified,liu2022certifiably,wicker2023robust}.

\paragraph{Human stakeholders in the context of AdvXAI} \citet{poursabzi2021manipulating} highlight the pivotal issue that human stakeholders are susceptible to ``information overload'' in the context of interpretability, i.e. not always presenting more explanations is better~\citep{baniecki2023grammar}. Exploiting this particular failure mode of XAI methods creates new possibilities for attacking their end users~\citep{lakkaraju2019faithful,schneider2022deceptive}. In practice, whether a particular attack or defense is successful will differ between application domains, e.g. explaining financial decisions for regulatory audit~\citep{shrotri2022constraint,carmichael2023unfooling}, or selecting bio-markers for scientific discovery~\citep{jimenez2020drug}. There are emerging works considering the interactivity of XAI process \citep[e.g. see][and references given there]{baniecki2023grammar,slack2023explaining}, in which case attacks and defenses could exploit the conversational framework of human-model interaction.

\paragraph{Ethics, impact on society, and law concerning AdvXAI}
Finally, we need to take into account the broader impact adversarial research has on society. How does AdvXAI fit into regulations like AI Act~\citep{floridi2021european}, the four-fifths rule of fairness~\citep{watkins2022fourfifths}, or the right to explanation~\citep{krishna2023bridging}? For example, commercial stakeholders can perform ``XAI-washing'', i.e. imply that a product or service uses XAI to pass some legal requirements when in fact explanations are not authentic, e.g. maliciously crafted with adversarial methods. These questions are yet to be answered. For a more philosophical consideration on explanation robustness, we refer the reader to the argument by \citet{hancox2020robustness} concerning epistemic and ethical reasons for seeking objective explanations.

\section*{Acknowledgements}

We thank participants and anonymous reviewers at the ``IJCAI 2023 Workshop on XAI'' for their valuable feedback. We would like to acknowledge Xingyu Zhao, Satoshi Hara and Martin Pawelczyk for contributing to an ongoing collection of relevant papers available at \url{https://github.com/hbaniecki/adversarial-explainable-ai}. This work was financially supported by the Polish National Science Centre grant number 2021/43/O/ST6/00347.

\clearpage

\appendix

\section{List of abbreviations and proper names (with references)}\label{app:abbr}

\begin{itemize}
    \item Anchors: high-precision model-agnostic explanations \citep{ribeiro2018anchors}
    \item CAM: class activation mapping \citep{zhou2016learning}
    \item CNN: convolutional neural network \citep{lecun1998gradientbased}
    \item DT: decision tree \citep{breiman1984classification}
    \item EOdds: equalized odds \citep{hardt2016equality}
    \item EOpp: equal opportunity \citep{hardt2016equality}
    \item \texttt{fairlearn} software \citep{weerts2023fairlearn}
    \item FFV: faceted feature visualization \citep{goh2021multimodal}
    \item GBDT: gradient boosting decision tree \citep{friedman2001greedy}
    \item GBP: guided backpropagation \citep{springenberg2015striving}
    \item GCN: graph convolutional network \citep{kipf2017semisupervised}
    \item GI: gradient input \citep{shrikumar2017learning}
    \item Grad-CAM: gradient-weighted class activation mapping \citep{selvaraju2020gradcam}
    \item GS: grid saliency \citep{hoyer2019grid}
    \item IG: integrated gradients \citep{sundararajan2017axiomatic}
    \item IME: interactions-based method for explanation \citep{vstrumbelj2014explaining}
    \item KNN: k-nearest neighbours \citep[see][section 13.3]{hastie2001elements}
    \item LIME: local interpretable model-agnostic explanations \citep{ribeiro2016why}
    \item LR: logistic regression \citep[see][section 4.4]{hastie2001elements}
    \item LRP: layer-wise relevance propagation \citep{bach2015onpixelwise}
    \item MLP: multi-layer perceptron \citep[see][chapter 11]{hastie2001elements}
    \item MUSE: model understanding subspace explanations \citep{lakkaraju2019faithful}
    \item PDP: partial dependence plot \citep[introduced in][section 8.2]{friedman2001greedy}
    \item PE: predictive equality \citep{corbett2017algorithmic}
    \item Relevance-CAM: relevance-weighted class activation mapping \citep{lee2021relevancecam}
    \item RF: random forest \citep{breiman2001random}
    \item RTS: real time saliency \citep{dabkowski2017real}
    \item SG: simple gradient \citep{simonyan2013deep}
    \item SHAP: Shapley additive explanations \citep{lundberg2017unified}
    \item SP: statistical (demographic) parity \citep[see][section 4.1]{mehrabi2021survey}
    \item SVM: support vector machine \citep{boser1992training}
    \item TCAV: testing with concept activation vectors \citep{kim2018interpretability}
    \item ViT: vision transformer \citep{dosovitskiy2021transformers}
    \item XGBoost: extreme gradient boosting \citep{tianqi2016xgboost}
\end{itemize}

\hbadness=6000 
\bibliographystyle{elsarticle-num-names} 
\bibliography{references}

\end{document}